\documentclass[
]{ceurart}

\usepackage{listings}
\begin{document}

\copyrightyear{2026}
\copyrightclause{Copyright for this paper by its authors.
  Use permitted under Creative Commons License Attribution 4.0
  International (CC BY 4.0).}

\conference{Authors' preprint.}

\title{LinkML-Scala: a Robust, Fast, and Portable Implementation of LinkML}

\author[1,2]{Piotr Sowiński}[%
orcid=0000-0002-2543-9461,
email=piotr@neverblink.eu 
]
\cormark[1]
\address[1]{NeverBlink, ul. Wspólna 56, 00-684 Warsaw, Poland}
\address[2]{Warsaw University of Technology, Pl. Politechniki 1, 00-661 Warsaw, Poland}

\author[1]{Kacper Grzymkowski}[%
orcid=0009-0008-9227-8240,
email=kacper.grzymkowski@neverblink.eu,
]

\author[1]{Andriy Plokhotnyuk}[%
orcid=0009-0006-8321-2202,
email=andriy@neverblink.eu
]

\cortext[1]{Corresponding author.}

\begin{abstract}
LinkML is a unified framework for data and domain modeling that spans diverse formats and ecosystems: JSON, RDF, CSV, SQL, spreadsheets and more. However, until now it had only one fully-featured implementation, written in Python, which is limited in terms of performance, portability, and behavior consistency. This restricts LinkML's usability in settings such as real-time schema editing and enterprise server applications.
To address these issues, we present LinkML-Scala: a robust, fast, and portable implementation of LinkML that covers the metamodel, runtime support, schema derivation, and generators for JSON Schema, SHACL, RDFS, and Table Schema. Written in Scala 3, it runs in the browser (JavaScript transpilation), on the JVM, and as native binaries. LinkML-Scala is distributed as an in-browser playground, a CLI application, a GitHub CI Action, and JVM / JavaScript libraries.
In our benchmarks, it outperforms the Python implementation in every tested scenario, on average by 22.9--38.5x.
We consider LinkML-Scala an important contribution toward increasing LinkML's adoption and we outline a plan for further work to ensure implementation interoperability and stability.
\end{abstract}


\maketitle

\section{Introduction}

Integrating data across systems and organizations remains a persistent challenge. Data is described using a plethora of formats and modeling languages -- JSON Schema, SQL DDL, spreadsheets, and countless bespoke conventions -- each tied to a particular technology stack. However, these formats capture structure but not meaning: they offer little support for expressing what the data is about, linking records across datasets, or aligning terms with shared vocabularies. The resulting models are fragmented and locked to their original ecosystem, making them hard to reuse or interlink.

On the other hand, Semantic Web standards such as RDFS~\cite{rdfs}, OWL~\cite{owl}, and SHACL~\cite{shacl} support rich, machine-readable semantics and interlinking across datasets. However, this expressiveness applies mainly to data represented as RDF. The vast majority of data in practice is held in relational databases, JSON documents, and spreadsheets, not RDF.
As a result, the benefits the semantic standards offer remain out of reach for most of the data that could benefit from them.

LinkML~\cite{moxon2025linkml} addresses this gap: it is a unified framework for data and domain modeling that works across different data formats and ecosystems. LinkML is designed to be easy to use and expressive enough to cover diverse types of data, from spreadsheets to JSONs and complex, interlinked RDF graphs. LinkML models are written in a YAML syntax, and then translated to other languages (e.g., JSON Schema, RDFS, SHACL, Table Schema~\cite{fowler2017frictionless}, SQL DDL, classes in OOP languages) with the use of \emph{generators}. This generated code can then be reused with tooling specific to the target ecosystem -- e.g., one can write a LinkML model, generate SHACL from it, and apply it with any SHACL validator.

So far, there was only one implementation of LinkML that covered the full modeling lifecycle (authoring, validation, generation): the \texttt{linkml} Python package. Over the course of our work, we found that this implementation has limitations that inhibit LinkML's usefulness in enterprise use cases:

\begin{enumerate}
    \item \textbf{Poor performance.} The implementation was never designed to be fast and the Python runtime is also limiting in this regard. This makes working with large models challenging and prohibits certain use cases entirely, such as real-time model linting in code editors.
    \item \textbf{Lack of portability.} Due to being tied to the Python ecosystem, \texttt{linkml} cannot run in browsers, native binaries, or the Java Virtual Machine, without extensive and fragile workarounds. This limits LinkML's usability in user interfaces, server applications, and edge devices.
    \item \textbf{Inconsistent behaviors and low reliability.} Code generators in \texttt{linkml} have many undocumented or off-spec behaviors and inconsistencies. A LinkML model may work with JSON Schema, but break unexpectedly in SHACL. This contributes to a poor user experience.
\end{enumerate}

To address these issues, we propose \textbf{LinkML-Scala}: a cross-platform, robust, and fast implementation of LinkML, covering the metamodel, runtime support, schema derivation, and generators for various languages (e.g., JSON Schema, SHACL, RDFS, Table Schema). LinkML-Scala aims to be a drop-in replacement for the \texttt{linkml} Python package, while being much faster, portable (spanning JavaScript, JVM, and native binaries), and more reliable.
Our contributions include: (1) an open-source implementation of LinkML-Scala; (2) tooling, including an in-browser playground, a CLI application, and a GitHub CI Action; (3) performance benchmarks comparing LinkML-Scala with the \texttt{linkml} Python package.

\section{LinkML-Scala}

LinkML-Scala is written in the Scala 3 language, which is key to its portability. Pure-Scala programs can run not only on the Java Virtual Machine, but can also also be transpiled to JavaScript with Scala.js~\cite{doeraene2013scala}, and compiled to native binaries with GraalVM~\cite{wimmer2021graalvm}.
Scala also offers excellent metaprogramming capabilities~\cite{stucki2020semantics,stucki2023scalable} and a very rich type system, which allow us to express very complex transformations succinctly and safely (with compile-time guarantees).
We followed the official \textit{Porting LinkML}\footnote{\url{https://linkml.io/linkml/howtos/port-linkml.html}} guide, starting with a Scala generator implemented in Python.
After LinkML-Scala reached self-hosting (Scala code generated from LinkML-Scala), we discarded the Python-to-Scala generator.

\textbf{Architecture.} LinkML-Scala has a modular design. At the core lies \emph{Runtime}, which contains the minimum components to support Scala classes generated from LinkML schemas.
To load schemas from YAML files into type-safe classes, we implemented a custom codec generator using the \texttt{scala-yaml} parser.
Custom Scala macros were needed to support LinkML's compact and simple dictionaries which  inject YAML dictionary keys into child object fields during decoding, a structural semantic that standard serialization libraries cannot handle.
The macros emit highly optimized and reflection-free decoding/encoding code at compile time while safely resolving recursive schema types. 
Scala's typeclass system (implicits)~\cite{oliveira2010type} allowed us to easily override edge cases by providing custom implicit codecs.

\emph{Metamodel} and \emph{SchemaView} packages implement LinkML's syntax and semantics.
The \emph{Metamodel} package contains Scala code generated from the LinkML metamodel using our own Scala generator.
The \emph{SchemaView} package is the toolkit for developing generators. 
It implements LinkML semantics, schema loading utilities, schema validation, and common utilities for generators.
All data classes are immutable, which facilitates easy caching using Scala's \texttt{lazy val}.
ElementView instances are used to encapsulate functionality specific to a type of a LinkML Element, such as text-to-meaning mappings for enums and slot derivation for classes.
Additional utilities provide functionalities such as inferring the inlining mode, case conversions, and mapping LinkML types to their runtime counterparts.
Many of these utilities use Scala enumerations or algebraic data types, which allows generators to use pattern matching to turn unhandled cases into compile-time errors.
This set-up allows us to move most logic out of generators and into the shared SchemaView, so writing consistent generators is much easier.

The \emph{Generators} package contains the generators that transform a LinkML schema into different output formats.
Currently, LinkML-Scala supports JSON Schema, SHACL, RDFS, Scala, Table Schema and LinkML (derivation) generators.
These generators take a variety of approaches, from direct string building (e.g., Scala generator), to intermediate AST generation with a decoupled serializer (e.g., JSON Schema, Table Schema).
New generators can be implemented easily as plugins, a pattern used extensively in NeverBlink's internal projects.

\textbf{Testing.} 
To ensure consistency between generators, LinkML-Scala hosts a model catalog for testing (dubbed the \emph{model zoo}), covering various edge cases of the LinkML specification.
These models are accompanied by instances in different serialization formats.
This allows testing whether the generators output schemas that accept valid instances and reject invalid instances, avoiding assertions on the structure of generator output, and only asserting its expected behavior.
We find this property useful for integration testing with languages that have their own validators, such as SHACL and JSON Schema.
Thanks to this, we are able to enforce a common data model for all generators.
The model catalog is language-agnostic and can be reused by other implementations of LinkML.

\section{Tooling and User Experience}

To facilitate uptake, LinkML-Scala is made available as: (1) a CLI application; (2) an in-browser playground; (3) a GitHub CI Action; (4) a library, published to Maven Central (JVM) and npm (JS). These artifacts were designed with user experience in mind -- we highlight some of these considerations below.

\textbf{CLI.} The \texttt{linkml-scala} command-line application currently allows one to lint/validate models (\texttt{linkml-scala validate}) and use generators (e.g., \texttt{linkml-scala generate shacl}). The tool is a single, self-contained native binary, compiled for Linux, macOS, and Windows. It can be easily installed via an installation script or the \texttt{mise} environment manager. The binary is compiled with GraalVM Native Image~\cite{wimmer2021graalvm}, which allows for near-instant startup times ($\sim$2.5ms to show the help text on Linux), in contrast to the typical 1--2s needed to start a JVM or a Python application. 
This greatly speeds up various scripted workloads (e.g., CI pipelines) and makes for a more responsive user experience.

\begin{figure}[htb]
    \centering
    \includegraphics[width=0.9\linewidth]{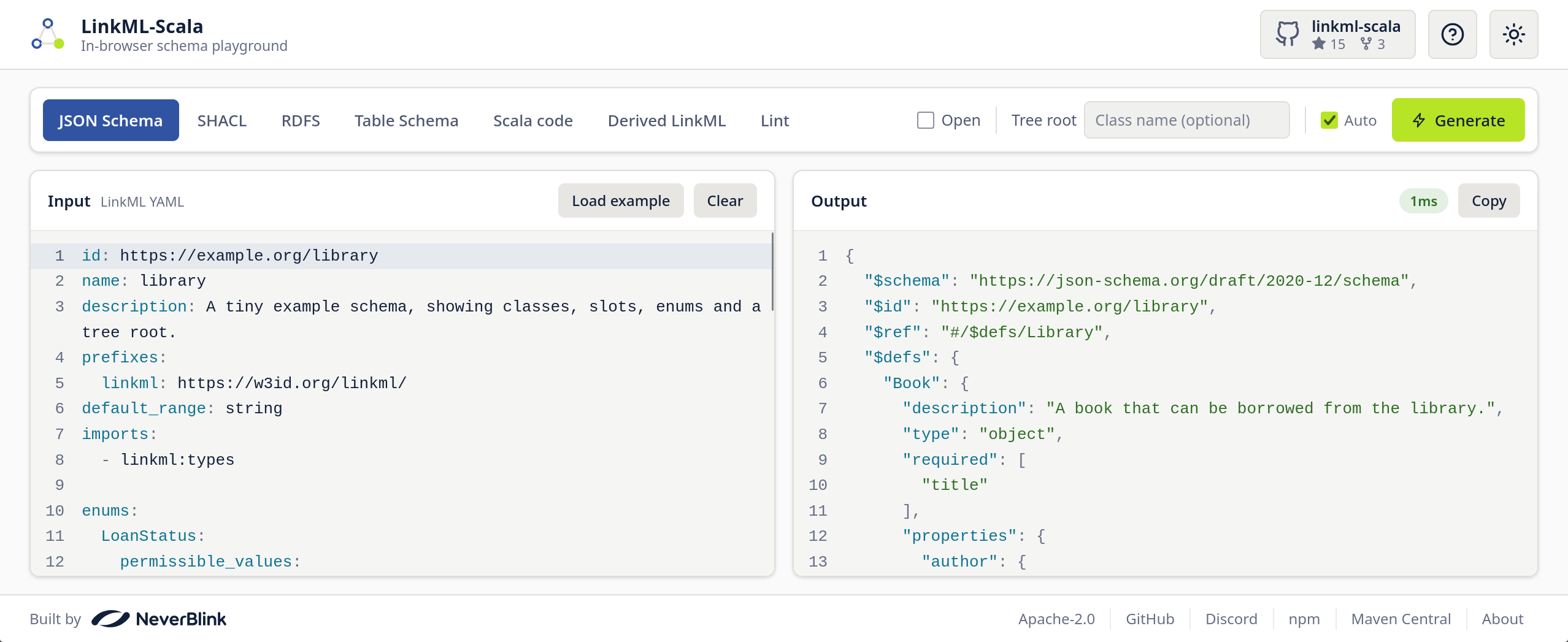}
    \caption{LinkML-Scala in-browser playground: \url{https://linkml.neverblink.eu/playground/}}
    \label{fig:playground}
\end{figure}

\textbf{Online playground.} Exploiting the fact that LinkML-Scala can compile to JavaScript, we built an in-browser playground\footnote{\url{https://linkml.neverblink.eu/playground/}} (Figure~\ref{fig:playground}). It allows one to edit and validate LinkML models, as well as use the built-in generators (e.g., live LinkML-to-SHACL conversion). In our view, it is a useful tool to help popularize LinkML, as it requires no installation or backend and can be used for quick experimentation.

\textbf{GitHub CI Action.} We package a ready-made GitHub CI Action\footnote{\url{https://github.com/marketplace/actions/linkml-linkml-scala}} that allows for fast schema validation and generation. The Action is written in pure JavaScript, so it runs on all platforms.

\textbf{Libraries.} We publish library JARs to Maven Central -- this allows for accessing the \texttt{SchemaView} interface, generators, and other features in Scala, Java, or other JVM languages. The library also allows for writing custom generators, a feature used extensively at NeverBlink for internal projects. 
We also publish an ES module to npm that exposes an API for running generators programmatically from JavaScript\footnote{\url{https://www.npmjs.com/package/@neverblink/linkml}}. The module includes TypeScript bindings to speed up application development.

\textbf{Error handling.} LinkML-Scala detects model issues at \texttt{SchemaView} level, not in generators, making it behave uniformly across generators. All errors point to a specific location in the model (line/column or JSON path) and have standardized messages. This speeds up model debugging and development.

\section{Evaluation}

We evaluated the performance of the two implementations by measuring the throughput of SHACL and JSON Schema generators (generations per second). The benchmarks were performed in two scenarios that emulate different usage patterns: \textbf{(1) Cold start}: end-to-end generation from CLI, emulating a workflow typical for CI scripts and schema authoring; \textbf{(2) Warm}: generator runs in a loop inside an interpreter/JIT, emulating a persistent server application.

\textbf{Benchmark setup.} We used LinkML-Scala 0.9.3 and LinkML (Python) 1.11.1. Test bench: Intel Core Ultra 9 285K CPU (3.7GHz, boost 5.7GHz), RAM 64GB DDR5-6400, Ubuntu Desktop 24.04 (Linux 6.14). Unless otherwise stated, we used OpenJDK 25.0.2+10-LTS and CPython 3.14.6 as the runtimes.

\textbf{Datasets.} We collected 11 diverse LinkML schemas for benchmarking by browsing the LinkML Schema Registry\footnote{\url{https://linkml.io/linkml-registry/registry/}} and publicly available GitHub repositories.
These schemas cover domains such as: cybersecurity (d3fend, iso27001), finance (cdm), biology (nmdc, crdch, include), and energy (tc57cim). Table~\ref{tab:datasets} presents a summary of all datasets.
Several schemas required minor manual repairs, such as fixing missing prefixes or invalid URIs.
We published the complete benchmark dataset on GitHub\footnote{\url{https://github.com/NeverBlink-labs/linkml-benchmark-schemas}}.

\begin{table}[htb]
  \centering
  \footnotesize
  \setlength{\tabcolsep}{4pt}
  \caption{Benchmark dataset statistics. \emph{Classes} and \emph{Attributes} are the materialized (induced) totals.}
  \label{tab:datasets}
  \begin{tabular}{lrrrrrrrrrrrr}
    \toprule
     & bridge2ai & cdm & chem-dcat & crdch & d3fend & fluxnova & include & iso27001 & nmdc & sssom & tc57cim \\
    \midrule
    Files & 1 & 37 & 5 & 1 & 1 & 17 & 1 & 1 & 14 & 1 & 1 \\
    Size (KiB) & 39.5 & 2237.9 & 117.4 & 1064.8 & 2591.3 & 244.3 & 56.3 & 251.7 & 558.8 & 59.4 & 2952.3 \\
    Classes & 38 & 779 & 89 & 41 & 4366 & 258 & 10 & 35 & 80 & 9 & 1528 \\
    Attributes & 160 & 3245 & 889 & 335 & 5250 & 2765 & 149 & 781 & 1655 & 110 & 34172 \\
    \bottomrule
  \end{tabular}
\end{table}

\begin{figure}[htb]
    \centering
    \includegraphics[width=1\linewidth]{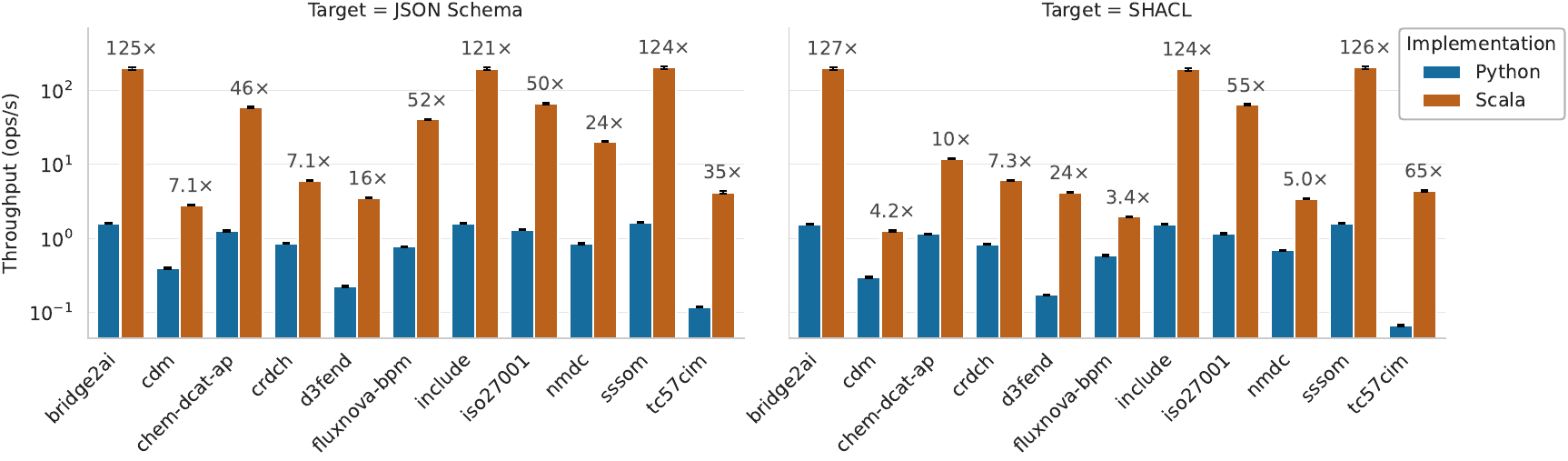}
    \caption{Cold-start benchmark results.}
    \label{fig:cold}
\end{figure}

\textbf{Cold-start benchmarks.} The benchmark is implemented as a shell script calling the hyperfine CLI tool~\cite{peter2023hyperfine}.
We used the CLI binary compiled with Oracle GraalVM Native Image 25.0.1+8.1 for LinkML-Scala, and the default LinkML (Python) CLI script. Results are presented in Figure~\ref{fig:cold}.

\begin{figure}[htb]
    \centering
    \includegraphics[width=1\linewidth]{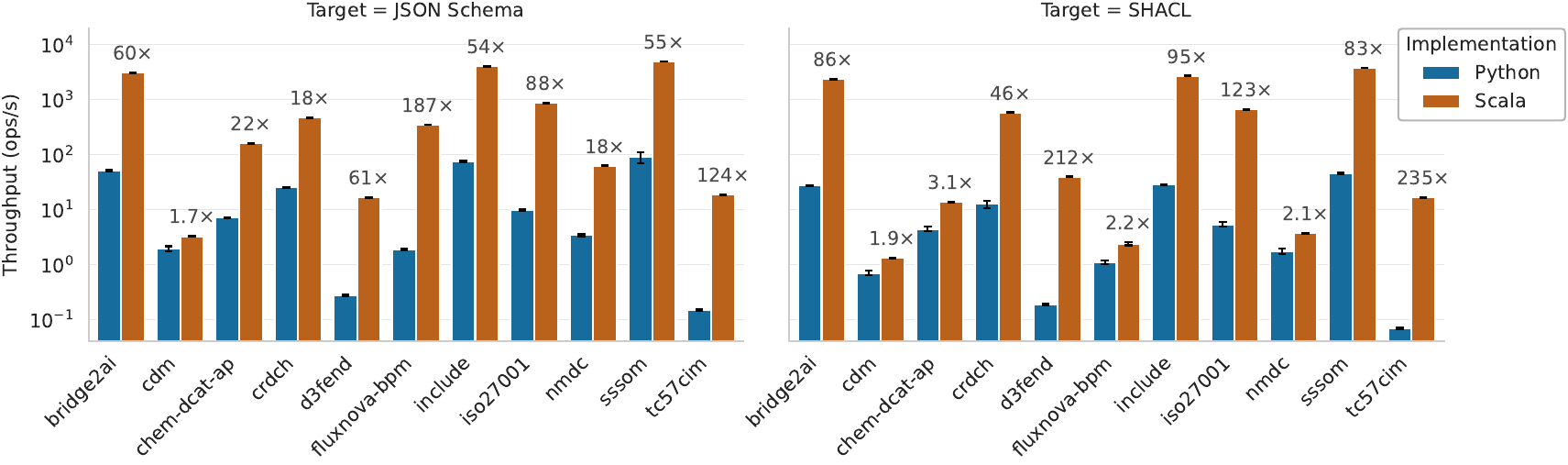}
    \caption{Warm benchmark results.}
    \label{fig:warm}
\end{figure}

\textbf{Warm benchmarks.} Benchmark has split implementations: LinkML-Scala uses the Java Microbenchmark Harness~\cite{web:jmh}, LinkML-Python uses a plain Python script.
Both implementations are allowed to parse schemas and resolve imports in their \emph{set-up} phase, meaning that all required models are loaded into memory before measurement.
Each warm scenario has 5 warm-up runs and 10 measure runs across 5 forks. Results are presented in Figure~\ref{fig:warm}.

\textbf{Results summary.} We calculate \emph{speedup} as the ratio between LinkML-Scala's throughput and LinkML (Python) throughput. The geometric mean of the speedup for the cold-start scenario is \textbf{36.5x} (JSON Schema) and \textbf{22.9x} (SHACL). For the warm scenario, the average speedup is \textbf{38.5x} (JSON Schema) and \textbf{26.8x} (SHACL). LinkML-Scala is faster than LinkML (Python) in every tested scenario.

\textbf{Discussion and limitations.} The speedup is not uniform across datasets. In warm JSON Schema benchmarks it ranges from 1.7x to 187x (two orders of magnitude). These differences may be caused by uneven feature coverage between the implementations, unoptimized code paths in LinkML-Scala, or very text-heavy schemas -- we will investigate these outliers in follow-up work. In cold-start benchmarks, for small datasets LinkML-Scala was mainly limited by the startup time of its GraalVM-compiled binary. We estimate that on a modern Linux machine \texttt{linkml-scala} starts in $\sim$2.5ms. Some datasets (e.g., \texttt{sssom}) were generated in only $\sim$5ms, making startup roughly half of the elapsed time.

\section{Conclusion and Future Work}

LinkML-Scala is a robust, fast, and portable implementation of LinkML, opening new application areas for this data modeling framework, such as responsive web editors and enterprise-grade server applications. 
It can be reused in a variety of ways, including a CLI application, programming libraries (JS and JVM), an online playground, and a GitHub Actions workflow. We also contribute the model catalog (test cases) and benchmark datasets, to be reused by the community.

In the immediate future, we are considering adding support for SQL DDL, Avro, Parquet, and Protobuf. We also plan to ingrate LinkML-Scala with Jelly~\cite{sowinski2025jelly}, to speed up processing of large SHACL/RDFS schemas.
We hope that LinkML-Scala will mark a turn in LinkML's development, as the second fully-featured implementation, encouraging the improvement of the LinkML specification as a shared interoperability basis. To that end, in the coming months we will contribute fixes and improvements to the specification. We also plan to engage directly with LinkML (Python) maintainers to improve the coverage and consistency of both implementations. We are also eager to see how the community will choose to use LinkML-Scala and what features should be developed next. \textbf{We invite feature requests on our issue tracker:} \url{https://github.com/NeverBlink-OSS/linkml-scala/issues}

\vspace{0.3cm}
\noindent
\textbf{Online playground:} \url{https://linkml.neverblink.eu/playground/} \\
\textbf{Code and documentation:} \url{https://github.com/NeverBlink-OSS/linkml-scala} \\
\textbf{Benchmark code and datasets:} \url{https://github.com/NeverBlink-labs/linkml-benchmark-schemas}

\vspace{0.3cm}
\textbf{Acknowledgements.} This work has been supported by the HEDGE-IoT project grant number 101136216 funded by the European Commission as part of the Horizon Europe Framework Programme. 

\textbf{Declaration on Generative AI.} During the preparation of this work, the authors used Anthropic Claude to assist in writing the benchmark code, data analysis scripts, and plotting scripts, and for drafting selected parts of the paper. After using this tool, the authors reviewed and edited the content, and performed manual cross-checks to ensure output validity. The authors take full responsibility for the publication's content.

\bibliography{bibliography}

\end{document}